\documentclass[aps,prd,twocolumn,nofootinbib]{revtex4-1}
\usepackage{epsfig}
\usepackage[colorlinks,linkcolor=blue,anchorcolor=blue,citecolor=blue,urlcolor=blue,breaklinks=true]{hyperref}
\usepackage{graphics}
\usepackage{slashed}
\usepackage{color}
\usepackage{amsmath}
\usepackage{bm}
\usepackage{setspace}

\begin{document}
\author{Ya-Peng Zhao$^{1}$}\email{zhaoyapeng2013@hotmail.com}
\author{Rui-Rui Zhang$^{1}$}
\author{Han Zhang$^{1,2}$}\email{zhanghan@nju.edu.cn}
\author{Hong-Shi Zong$^{1,3,4}$}\email{zonghs@nju.edu.cn}
\address{$^{1}$ Department of Physics, Nanjing University, Nanjing 210093, China}
\address{$^{2}$Collaborative Innovation Center of Advanced Microstructures, Nanjing University, Nanjing 210093, China}
\address{$^{3}$State Key Laboratory of Theoretical Physics, Institute of Theoretical Physics, CAS, Beijing, 100190, China}
\address{$^{4}$Joint Center for Particle, Nuclear Physics and Cosmology, Nanjing, 210093, China}
\title{Chiral phase transition from the Dyson-Schwinger equations in a finite spherical volume}

\begin{abstract}
 Within the framework of Dyson-Schwinger equations and by means of Multiple Reflection Expansion, we study the finite volume effects on the chiral phase transition in a sphere, especially discuss its influence on the location of the possible critical end point (CEP).
According to our calculations, when we take the sphere instead of cube as a research, the influence of finite volume effects on phase transition is not as significant as previously calculated.
 For instance, as the radius of spherical volume decreases from infinite to $2 \mathrm{fm}$, at zero chemical potential and finite temperature, the critical temperature $T_{c}$ has only a slight drop. And at finite chemical potential and finite temperature, the location of CEP shifts toward smaller temperature and higher chemical potential, but the amplitude of variation does not exceed $20\%$.
 So we find that not only the size of the volume, but also the shape of the volume will have a considerable impact on the phase transition.
\bigskip

\bigskip

\noindent PACS Number(s): 12.38.Mh, 11.10.Wx, 64.60.an

\end{abstract}


\maketitle

\section{INTRODUCTION}
It is widely believed that with the increases of temperature or/and chemical potential, the strongly-interacting matter undergoes a phase transition from hadronic matter to quark-gluon plasma (QGP), which now can be reproduced by relativistic heavy-ion collisions (RHIC) at CERN (France/Switzerland), BNL (USA), and GSI (Germany)~\cite{ADAMS2005102,SHURYAK200948}. Theoretically, on the one hand, ab initio lattice QCD~\cite{Bors¨¢nyi2010} simulation found the transition is a crossover at low chemical potential. On the other hand, QCD effective model~\cite{Lu2015,doi:10.1142/S0217751X15501997} calculations generally indicate that the phase transition is a first order at high chemical potential, and in the middle of the chemical potential, there exist a critical end point (CEP) where the first order phase transition ends. One important goal in the RHIC is to determine the existence and the location of the CEP. For this purpose, the second phase of the beam energy scan at RHIC will be performed between 2019 and 2021~\cite{LUO201675}.

It should be noted that many previous calculations about the location of CEP are based on infinite thermodynamics systems. However, the QGP system produced in RHIC has a finite volume undoubtedly. The volume of homogeneity before freeze-out for Au-Au and Pb-Pb collisions ranges between approximately $50\sim250$ $\mathrm{fm}^{3}$~\cite{PhysRevC.85.044901} based on the UrQMD transport approach~\cite{BASS1998255}. And the smallest quark-gluon plasma (QGP) system produced at RHIC could be as low as $(2\ \mathrm{fm})^{3}$ as the Ref.~\cite{JPhysG.38.085101} estimated. Therefore, theoretically studying whether the finite volume effects has a significant impact on CEP is important for RHIC experiment and it has been studied within the Nambu-Jona-Lasinio (NJL) model~\cite{doi:10.1142/S0217732318502322}, Polyakov-Nambu-Jona-Lasinio (PNJL) model~\cite{IntJModPhysA.32.1750067,PhysRevD.87.054009}, quark-meson model~\cite{PhysRevD.90.054012} and Dyson-Schwinger equations (DSEs)~\cite{Shi:2018tsq,LI2019298,1674-1137-42-2-023101}.

It is well known that when the volume of the strongly interacting system we studied is small enough, not only its size but also its shape will have an important impact on the QCD phase transition. However, it should be noted that in most of the previous calculations, for the sake of convenience, people usually use cube to simulate the fireball produced in RHIC, which ignores the influence of different shapes on the phase transition. Therefore, in order to get closer to the shape of the fireball produced by the RHIC experiment, we use the sphere and by means of the MRE~\cite{BALIAN1970401} to study the finite volume phase transition in the framework of DSEs.
Compared to other effective models, the DSEs take the quarks and gluons as the fundamental degrees of freedom, have both confinement and dynamical chiral symmetry breaking (DCSB) effects. And it has provided many insights into the QCD phase diagram, for instance, the chiral and deconfinement phase transition~\cite{FISCHER20131036,PhysRevD.93.036006,PhysRevD.91.056003}.

This paper is organized as follows: In Sec. II, we give a brief introduction to the quark gap equation in a finite spherical volume at finite temperature and finite chemical potential. In Sec. III, we study the finite volume effects on the chiral phase transition, especially its influence on the behaviour of the CEP. Finally, we will give a brief summary in Sec. IV.
\section{quark gap equation in a finite spherical volume}\label{ok}
The DSEs is a suitable QCD-connected non-perturbative method and it is widely used to study hadron physics~\cite{PhysRevD.92.114034,PhysRevD.93.114033} and QCD phase diagram. At zero temperature and zero chemical potential, the DSE of the quark propagator, namely the quark gap equation reads~\cite{1674-1137-42-2-023101}~\footnote{Here we work in Euclidean space, take the $N_{f}=2$ and $N_{c}=3$. Moreover, renormalization is actually unnecessary because of we employ a ultra-violet finite model.}
\begin{eqnarray}
S(p)^{-1}=S_{0}(p)^{-1}+\frac{4}{3}\int\frac{d^{4}q}{(2\pi)^4}g^{2}D_{\mu\nu}(p-q)\gamma_{\mu}S(q)\Gamma_{\nu},\nonumber\\
\end{eqnarray}
where $S(p)^{-1}$ is the inverse of the dressed quark propagator and $S_{0}(p)^{-1}$ is the inverse of the free one. $g$ is the coupling constant of strong interaction, $D_{\mu\nu}(p-q)$ is the dressed gluon propagator, and $\Gamma_{\nu}$ is the one-particle-irreducible quark-gluon vertex.
According to the Lorentz structure analysis, $S(p)^{-1}$ can generally be decomposed as~\cite{ROBERTS2000S1}
\begin{eqnarray}
S(p)^{-1}=i{\not\!p}A(p^{2})+B(p^{2}),
\end{eqnarray}
where $A(p^{2})$ and $B(p^{2})$ are scalar functions of $p^{2}$. Note that for the free quark propagator $S_{0}(p)^{-1}$, scalar functions $A=1, B=m$.

Next we extend the quark gap equation to finite $T$ and $\mu$ and it reads
\begin{eqnarray}
S(\vec{p},\tilde{\omega}_{n})^{-1}&=&S_{0}(\vec{p},\tilde{\omega}_{n})^{-1}+\frac{4}{3}T\int\!\!\!\!\!\!\!\!\sum g^{2}\nonumber\\
&&\times D_{\mu\nu}(\vec{k},\Omega_{nl})\gamma_{\mu}S(\vec{q},\tilde{\omega}_{l})\Gamma_{\nu},
\end{eqnarray}
where
\begin{eqnarray}
S_{0}(\vec{p},\tilde{\omega}_{n})^{-1}=i\vec{\gamma}\cdot \vec{p}+i\gamma_{4}\tilde{\omega}_{n}+m,
\end{eqnarray}
$\vec{k}=\vec{p}-\vec{q}$, $\Omega_{nl}=\omega_{n}-\omega_{l}$, $\tilde{\omega}_{n}=\omega_{n}+i\mu$, $\omega_{n}=(2n+1)\pi T, n\in \mathbf{Z}$ and\ \ $\int\!\!\!\!\!\!\!\!\sum$ denotes $\sum_{l}\int\frac{d^{3}\vec{q}}{(2\pi)^3}$.
Nevertheless, due to the breaking of $O(4)$ symmetry down to $O(3)$ symmetry, the Lorentz structure of $S(\vec{p},\tilde{\omega}_{n})^{-1}$ now have to be decomposed as
\begin{eqnarray}
S(\vec{p},\tilde{\omega}_{n})^{-1}&=&i{\not\!\vec{p}}A(\vec{p},\tilde{\omega}_{n})+\mathbf{1}B(\vec{p},\tilde{\omega}_{n})\nonumber\\
&&+i\gamma_{4}\tilde{\omega}_{n}C(\vec{p},\tilde{\omega}_{n})+{\not\!\vec{p}}\gamma_{4}\tilde{\omega}_{n}D(\vec{p},\tilde{\omega}_{n})
\end{eqnarray}
where ${\not\!\vec{p}}=\vec{\gamma}\cdot\vec{p}$, $\vec{\gamma}=(\gamma_{1}, \gamma_{2}, \gamma_{3})$, and the four scalar functions $F=A, B, C, D$ are complex and satisfy the condition
\begin{eqnarray}
F(\vec{p},\tilde{\omega}_{n})^{\ast}=F(\vec{p},\tilde{\omega}_{-n-1}),
\end{eqnarray}
which can be used to verify the accuracy of your numerical calculations.
In the next calculations we ignore the function $D$ because it is power-law suppressed in the ultra-violate region. And at zero $T$ but finite $\mu$, $D$ vanishes exactly as~\cite{Rusnak1995} shows because the corresponding tensor structure has the wrong transformation properties under time reversal. So, the widely used structure of $S(\vec{p},\tilde{\omega}_{n})^{-1}$ as follows
\begin{eqnarray}\label{propagator}
S(\vec{p},\tilde{\omega}_{n})^{-1}&=&i{\not\!\vec{p}}A(\vec{p},\tilde{\omega}_{n})+\mathbf{1}B(\vec{p},\tilde{\omega}_{n})\nonumber\\
&&+i\gamma_{4}\tilde{\omega}_{n}C(\vec{p},\tilde{\omega}_{n}),
\end{eqnarray}

Now, we are ready to introduce the quark gap equation in a finite spherical volume, and for taking the finite volume effects into account we consider the MRE formalism~\cite{PhysRevD.67.085010,PhysRevD.72.054009,PhysRevD.50.3328} which modifies the density of states as follows
\begin{eqnarray}
\rho_{MRE}(p,m,R)=1+\frac{6\pi^{2}}{pR}f_{S}+\frac{12\pi^{2}}{(pR)^{2}}f_{C},
\end{eqnarray}
where $f_{S}$ denote the surface contribution to the density of states
\begin{eqnarray}
f_{S}=-\frac{1}{8\pi}(1-\frac{2}{\pi}\mathrm{arctan}\frac{p}{m}),
\end{eqnarray}
and the curvature contribution is given by Madsen$^{,}$s ansatz~\cite{PhysRevD.50.3328}
\begin{eqnarray}
f_{C}=\frac{1}{12\pi^{2}}[1-\frac{3p}{2m}(\frac{\pi}{2}-\mathrm{arctan}\frac{p}{m})],
\end{eqnarray}
which takes the finite quark mass contribution into account. It should be noted that there have different interpretations of $m$ in the MRE formula when it is applied to nonperturbative calculations, for example, (P)NJL model, see Refs.~\cite{PhysRevD.67.085010,PhysRevD.72.054009,PhysRevC.88.045803}. Here, in this paper, we treat $m$ as current quark mass as Refs.~\cite{PhysRevD.72.054009,PhysRevC.88.045803} instead of constitute quark mass~\cite{PhysRevD.67.085010}.

For $m\neq 0$, the main problem with the MRE is that it predict a negative density of states at samll $p$ where in reality there are no states~\cite{PhysRevD.60.054011}. Therefore we remove it by introducing an IR cutoff ($\Lambda_{IR}$) in momentum space as Refs.~\cite{PhysRevD.72.054009,PhysRevC.88.045803}. Actually for anti-periodic boundary conditions of spatial directions, in a cubic box of size $L$, we have $\vec{p}^{2}=\frac{4\pi^{2}}{L^{2}}\sum_{i=1}^{3}(n_{i}+\frac{1}{2})^{2}$, $n_{i}=0, \pm1, \pm2 \cdot\cdot\cdot$. The minimum momentum $|p_{min}|=\frac{\pi}{L}$, similar to $\Lambda_{IR}$. So in the next calculations, the following replacement must be performed.
\begin{eqnarray}
\int_{0}^{\Lambda,\infty}\frac{ d^{3}\vec{p}}{(2\pi)^3}\cdots\rightarrow\int_{\Lambda_{IR}}^{\Lambda,\infty}\frac{ d^{3}\vec{p}}{(2\pi)^3}\rho_{MRE}\cdots,
\end{eqnarray}
where the $\Lambda_{IR}$ is the largest solution of the equation $\rho_{MRE}(p,m,R)=0$ with respect to the momentum $p$.
Thus in a finite spherical volume, the quark gap equation becomes
\begin{eqnarray}\label{finite}
S(\vec{p},\tilde{\omega}_{n})_{F}^{-1}&=&S_{0}(\vec{p},\tilde{\omega}_{n})^{-1}+\frac{4}{3}T\int\!\!\!\!\!\!\!\!\sum_{f}g^{2}\rho_{MRE}\nonumber\\
&&\times D_{\mu\nu}(\vec{k},\Omega_{nl})\gamma_{\mu}S(\vec{q},\tilde{\omega}_{l})\Gamma_{\nu},
\end{eqnarray}
where now\ \ $\int\!\!\!\!\!\!\!\!\sum_{f}$ denotes $\sum_{l}\int_{\Lambda_{IR}}\frac{d^{3}\vec{p}}{(2\pi)^3}$.

To solve the quark gap equation, truncations are inevitably. Here we employ the Rainbow truncation~\cite{PhysRevD.94.076009,PhysRevD.94.094030}
\begin{eqnarray}\label{rainbow}
\Gamma_{\mu}(p_{n},q_{l})=\gamma_{\mu},
\end{eqnarray}
which is widely used in studies of hadron physics and QCD phase diagram.
We also employ the widely used gluon propagator model as Refs.~\cite{PhysRevC.60.055214,PhysRevC.56.3369,PhysRevLett.106.172301}, which has the form
\begin{eqnarray}\label{gluon}
g^{2}D_{\mu\nu}(k_{\Omega})&=&\mathcal{G}(k_{\Omega}^{2})(\delta_{\mu\nu}-k_{\Omega}^{\mu}k_{\Omega}^{\mu}/k_{\Omega}^{2}),
\end{eqnarray}
where
\begin{eqnarray}\label{G}
\mathcal{G}(k_{\Omega}^{2})=\frac{4\pi^{2}}{\omega^{6}}D_{0}k_{\Omega}^{2}e^{-k_{\Omega}^{2}/\omega^{2}},
\end{eqnarray}
and $k_{\Omega}=(\vec{k},\Omega_{nl})$, $\delta_{\mu\nu}$=diag\{+1,+1,+1,+1\}.

The related parameters, $D_{0}$ and $\omega$ are usually fixed by observables in hadron physics: the pion mass $m_{\pi}=0.139\ \mathrm{GeV}$ and the pion decay constant $f_{\pi}=0.095\ \mathrm{GeV}$. Here we use the typical values, that is $\omega=0.5\ \mathrm{GeV}, D_{0}=1.0\ \mathrm{GeV}^{2}$~\cite{LI2019298} and the current quark mass $m=0.005\ \mathrm{GeV}$.
\section{finite volume effects on the qcd chiral phase diagram}\label{bigtwo}
In this section, we study the chiral phase transition in a finite spherical volume, especially discuss its influence on the location of the CEP. We first solve the quark gap equation. The procedure is to insert Eqs. (\ref{propagator},\ref{rainbow},\ref{gluon},\ref{G}) into Eqs. (\ref{finite}), multiply each side by $-i{\not\!\vec{p}}$, $-i\gamma_{4}\tilde{\omega}_{n}$ and $\mathbf{1}_{4}$ respectively, and then take traces on both sides. Note that $S(\vec{p},\tilde{\omega}_{n})$ and $S(\vec{p},\tilde{\omega}_{n})^{-1}$ have the same Lorentz structure. The coupled non-linear equations about scalar functions $A, B, C$ can then be obtained as follows
\begin{figure}
\includegraphics[width=0.47\textwidth]{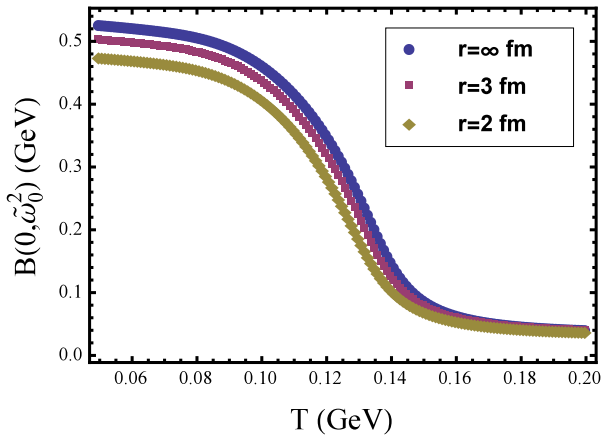}
\caption{Plot $B(0,\tilde{\omega}_{0}^{2})$ as a function of $T$ at $\mu=0$ for three different radius $r$.}
\label{Fig:bu0}
\end{figure}
\begin{figure}
\includegraphics[width=0.47\textwidth]{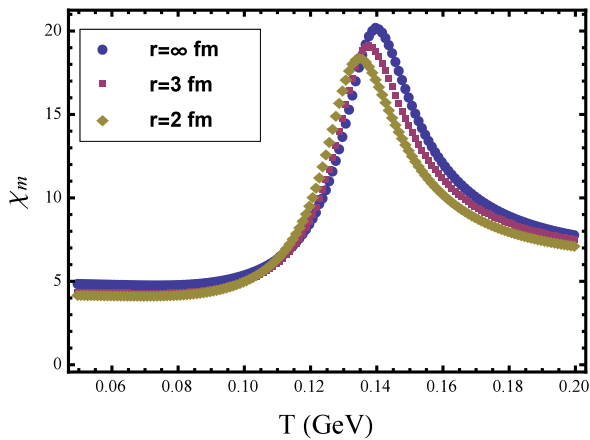}
\caption{Plot the chiral susceptibility $\chi_{m}$ as a function of $T$ at $\mu=0$ for three different radius $r$}
\label{Fig:kaim}
\end{figure}
\begin{figure}
\includegraphics[width=0.47\textwidth]{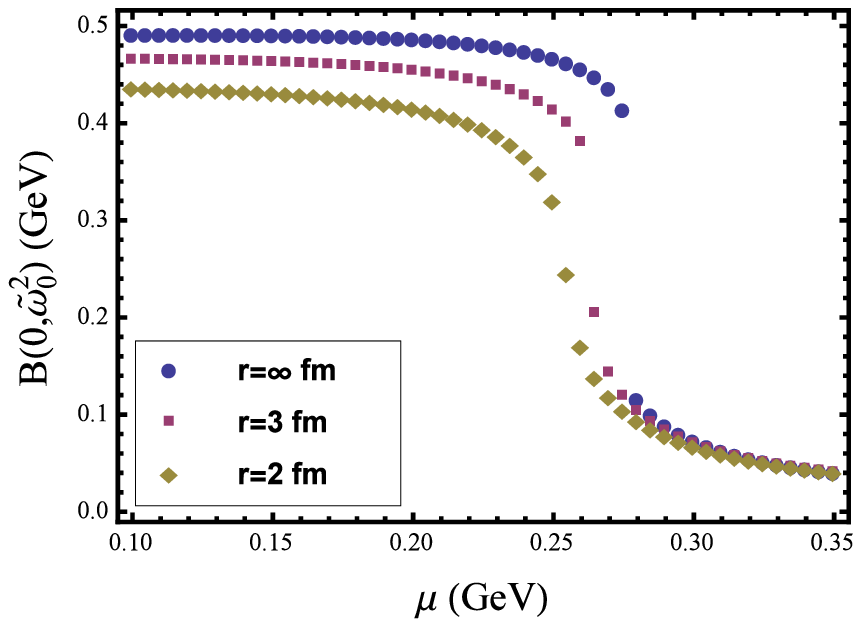}
\caption{Plot $B(0,\tilde{\omega}_{0}^{2})$ as a function of $\mu$ at $T=0.09\ \mathrm{GeV}$ for three different radius $r$.}
\label{Fig:bt90}
\end{figure}
\begin{figure}
\includegraphics[width=0.47\textwidth]{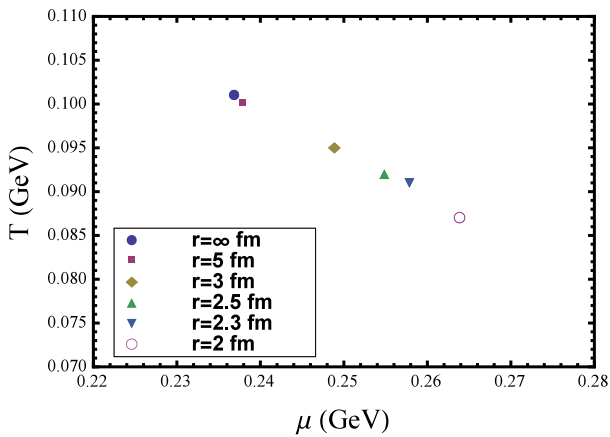}
\caption{Plot the volume dependence of the CEP.}
\label{Fig:cep}
\end{figure}
\begin{widetext}
\begin{eqnarray}
A_{pn}&=&1+\frac{4T}{3\vec{p}^{2}}\int\!\!\!\!\!\!\!\!\sum_{f}\rho_{MRE}\frac{\mathcal{G}(k_{\Omega}^{2})}{\vec{q}^{2}A^{2}_{ql}+\tilde{\omega}_{l}^{2}C^{2}_{ql}+B^{2}_{ql}}\{A_{ql}[\vec{p}\cdot\vec{q}+\frac{2(\vec{k}\cdot\vec{p})(\vec{k}\cdot\vec{q})}{k^{2}}]+C_{ql}\frac{2\tilde{\omega}_{l}\Omega_{nl}(\vec{k}\cdot\vec{p})}{k^{2}}\}
\end{eqnarray}
\begin{eqnarray}
C_{pn}&=&1+\frac{4T}{3\tilde{\omega}_{n}}\int\!\!\!\!\!\!\!\!\sum_{f}\rho_{MRE}\frac{\mathcal{G}(k_{\Omega}^{2})}{\vec{q}^{2}A^{2}_{ql}+\tilde{\omega}_{l}^{2}C^{2}_{ql}+B^{2}_{ql}}\{A_{ql}\frac{2\Omega_{nl}(\vec{k}\cdot\vec{q})}{k^{2}}+C_{ql}\tilde{\omega}_{l}(1+\frac{2\Omega_{nl}^{2}}{k^{2}})\}
\end{eqnarray}
\begin{eqnarray}
B_{pn}&=&m+\frac{4T}{3}\int\!\!\!\!\!\!\!\!\sum_{f}\rho_{MRE}\frac{3\mathcal{G}(k_{\Omega}^{2})B_{ql}}{\vec{q}^{2}A^{2}_{ql}+\tilde{\omega}_{l}^{2}C^{2}_{ql}+B^{2}_{ql}}
\end{eqnarray}
\end{widetext}
These coupled equations can be numerically solved by iteration.
In Fig. \ref{Fig:bu0}, we show $B(0,\tilde{\omega}_{0}^{2})$ as a function of $T$ for three different volumes. Firstly, $B(0,\tilde{\omega}_{0}^{2})$ decreases continuously with increasing temperature, which means the transition is a crossover.
Secondly, $B(0,\tilde{\omega}_{0})$ decreases as the volume decreases. For example, at $T=0.05\ \mathrm{GeV}$, we find that $B(0,\tilde{\omega}_{0})$ is reduced from $0.525\ \mathrm{GeV}$ to $0.473\ \mathrm{GeV}$ , which means the DCSB effect becomes weak. This is consist with other model calculations~\cite{PhysRevD.87.054009}.

The crossover behavior can be further studied by the chiral susceptibility which is defind as~\cite{PhysRevC.59.1751}
\begin{eqnarray}
\chi_{m}(T,\mu)=\frac{\partial B(0,\tilde{\omega}_{0}^{2})}{\partial m}
\end{eqnarray}
and the volume dependence are plotted in Fig. \ref{Fig:kaim}. We find that the critical temperature $T_{c}$ shows a slight volume dependence and it only decreases from $0.14\ \mathrm{GeV}$ to $0.135\ \mathrm{GeV}$. So, the conclusion is that at zero chemical potential and finite temperature, the influence of finite volume effects on chiral phase transition is not so obvious.
In Fig. \ref{Fig:bt90}, we plot the $B(0,\tilde{\omega}_{0}^{2})$ as a function of $\mu$ at $T=0.09\ \mathrm{GeV}$ for three different volumes. At infinite volume, $B(0,\tilde{\omega}_{0}^{2})$  appears a sudden discontinuity at a critical value $\mu_{c}=0.275\ \mathrm{GeV}$, which indicates the first order phase transition happens. But when the radius is reduced to $2\ \mathrm{fm}$, $B(0,\tilde{\omega}_{0}^{2})$ changes continuity and a crossover happens. This indicate that under $T=0.09\ \mathrm{GeV}$, the finite size effect has a significant impact on the phase diagram, that is, on the location of the CEP.

Therefore, we plot the volume dependence of the CEP in Fig. \ref{Fig:cep}. First of all, when the radius of spherical volume is larger than $5\ \mathrm{fm}$, the chiral phase diagram is almost the same as the infinite system. In the next place, as the radius of spherical volume decreases, we note that the CEP shifts toward smaller temperature and higher chemical potential.
Thus, for the CEP search in the RHIC experiments, according to our calculation, when the radius is below $3\ \mathrm{fm}$, the finite volume effects may not be negligible.
However, even if the radius is as small as $2\ \mathrm{fm}$, the location of the CEP shifts from $(\mu_{E},T_{E})=(0.237\ \mathrm{GeV},0.101\ \mathrm{GeV})$ to $(\mu_{E},T_{E})=(0.264\ \mathrm{GeV},0.087\ \mathrm{GeV})$, and the variation does not exceed $20\%$.
It is interesting here to compare our results with other DSE calculations~\cite{LI2019298,Shi:2018tsq}. The main difference is that when we study the sphere instead of cube, we find that the influence of finite volume effects on phase transitions is much weaker.
For instance, in Ref.~\cite{LI2019298}, the CEP moves from $(\mu_{E},T_{E})=(0.24\ \mathrm{GeV},0.10\ \mathrm{GeV})$ at $L=\infty$ to $(\mu_{E},T_{E})=(0.37\ \mathrm{GeV},0.025\ \mathrm{GeV})$ at $L=2.1 \ \mathrm{fm}$, and the variation more than $60\%$. In Ref.~\cite{Shi:2018tsq}, the variation more than $70\%$.
This reflects a fact that when we study the influence of finite volume effects on QCD phase transition, we should consider not only the size, but also the shape of the volume.
Finally, it is a little different from the PNJL model~\cite{PhysRevD.87.054009} in which the $\mu$ of CEP is almost a constant value.
\section{Summary and Conclusion}\label{summary}
Within the framework of DSEs, for the first time, we try to consider the influence of the finite volume effects on the chiral phase transition in a sphere. For taking the finite volume effects into account we consider the MRE formalism in which it also properly incorporate the surface and curvature effects. And it has been used to study thermodynamic quantities in the PNJL model~\cite{Grunfeld2018} and color superconducting in the NJL model~\cite{PhysRevD.72.054009}.
Our main conclusion is that not only the size of the volume, but also the shape of the volume will have a sizable impact on the phase transition.
Relatively speaking, we find that the influence of finite volume effects on chiral phase transitions in the sphere is not as significant as in the cube. For example, at zero chemical potential and finite temperature, the $T_{c}$ remains almost the same value as the volume decreases. At finite chemical potential and finite temperature, our results show that, as the radius of spherical volume decreases, the location of CEP shifts toward smaller temperature and higher chemical potential. But the amplitude of variation does not exceed $20\%$. Therefore, we should use a shape closer to the QGP fireball produced by RHIC in order to better study the finite volume phase transition.
Finally, in the next step, we intend to go beyond the bare vertex approximation~\cite{PhysRevD.22.2542,PhysRevLett.106.072001} and see if it will make a difference.

\acknowledgments
This work is supported by the National Natural Science Foundation of China (under Grants No. 11475085, No. 11535005, No. 11690030 and No. 11574145).

\bibliography{CPC}

\begin{thebibliography}{38}%
\makeatletter
\providecommand \@ifxundefined [1]{%
 \@ifx{#1\undefined}
}%
\providecommand \@ifnum [1]{%
 \ifnum #1\expandafter \@firstoftwo
 \else \expandafter \@secondoftwo
 \fi
}%
\providecommand \@ifx [1]{%
 \ifx #1\expandafter \@firstoftwo
 \else \expandafter \@secondoftwo
 \fi
}%
\providecommand \natexlab [1]{#1}%
\providecommand \enquote  [1]{``#1''}%
\providecommand \bibnamefont  [1]{#1}%
\providecommand \bibfnamefont [1]{#1}%
\providecommand \citenamefont [1]{#1}%
\providecommand \href@noop [0]{\@secondoftwo}%
\providecommand \href [0]{\begingroup \@sanitize@url \@href}%
\providecommand \@href[1]{\@@startlink{#1}\@@href}%
\providecommand \@@href[1]{\endgroup#1\@@endlink}%
\providecommand \@sanitize@url [0]{\catcode `\\12\catcode `\$12\catcode
  `\&12\catcode `\#12\catcode `\^12\catcode `\_12\catcode `\%12\relax}%
\providecommand \@@startlink[1]{}%
\providecommand \@@endlink[0]{}%
\providecommand \url  [0]{\begingroup\@sanitize@url \@url }%
\providecommand \@url [1]{\endgroup\@href {#1}{\urlprefix }}%
\providecommand \urlprefix  [0]{URL }%
\providecommand \Eprint [0]{\href }%
\providecommand \doibase [0]{http://dx.doi.org/}%
\providecommand \selectlanguage [0]{\@gobble}%
\providecommand \bibinfo  [0]{\@secondoftwo}%
\providecommand \bibfield  [0]{\@secondoftwo}%
\providecommand \translation [1]{[#1]}%
\providecommand \BibitemOpen [0]{}%
\providecommand \bibitemStop [0]{}%
\providecommand \bibitemNoStop [0]{.\EOS\space}%
\providecommand \EOS [0]{\spacefactor3000\relax}%
\providecommand \BibitemShut  [1]{\csname bibitem#1\endcsname}%
\let\auto@bib@innerbib\@empty
\bibitem [{\citenamefont {Adams}\ and\ \citenamefont {et~al.
  (STAR~Collaboration)}(2005)}]{ADAMS2005102}%
  \BibitemOpen
  \bibfield  {author} {\bibinfo {author} {\bibfnamefont {J.}~\bibnamefont
  {Adams}}\ and\ \bibinfo {author} {\bibnamefont {et~al.
  (STAR~Collaboration)}},\ }\href {\doibase
  https://doi.org/10.1016/j.nuclphysa.2005.03.085} {\bibfield  {journal}
  {\bibinfo  {journal} {Nucl. Phys. A}\ }\textbf {\bibinfo {volume} {757}},\
  \bibinfo {pages} {102 } (\bibinfo {year} {2005})},\ \bibinfo {note} {first
  Three Years of Operation of RHIC}\BibitemShut {NoStop}%
\bibitem [{\citenamefont {Shuryak}(2009)}]{SHURYAK200948}%
  \BibitemOpen
  \bibfield  {author} {\bibinfo {author} {\bibfnamefont {E.}~\bibnamefont
  {Shuryak}},\ }\href {\doibase https://doi.org/10.1016/j.ppnp.2008.09.001}
  {\bibfield  {journal} {\bibinfo  {journal} {Prog. Part. Nucl. Phys}\ }\textbf
  {\bibinfo {volume} {62}},\ \bibinfo {pages} {48 } (\bibinfo {year}
  {2009})}\BibitemShut {NoStop}%
\bibitem [{\citenamefont {Bors{\'a}nyi}\ \emph {et~al.}(2010)\citenamefont
  {Bors{\'a}nyi}, \citenamefont {Fodor}, \citenamefont {Hoelbling},
  \citenamefont {Katz}, \citenamefont {Krieg}, \citenamefont {Ratti},\ and\
  \citenamefont {Szab{\'o}}}]{Bors¨¢nyi2010}%
  \BibitemOpen
  \bibfield  {author} {\bibinfo {author} {\bibfnamefont {S.}~\bibnamefont
  {Bors{\'a}nyi}}, \bibinfo {author} {\bibfnamefont {Z.}~\bibnamefont {Fodor}},
  \bibinfo {author} {\bibfnamefont {C.}~\bibnamefont {Hoelbling}}, \bibinfo
  {author} {\bibfnamefont {S.~D.}\ \bibnamefont {Katz}}, \bibinfo {author}
  {\bibfnamefont {S.}~\bibnamefont {Krieg}}, \bibinfo {author} {\bibfnamefont
  {C.}~\bibnamefont {Ratti}}, \ and\ \bibinfo {author} {\bibfnamefont {K.~K.}\
  \bibnamefont {Szab{\'o}}},\ }\href {\doibase 10.1007/JHEP09(2010)073}
  {\bibfield  {journal} {\bibinfo  {journal} {JHEP}\ }\textbf {\bibinfo
  {volume} {2010}},\ \bibinfo {pages} {73} (\bibinfo {year}
  {2010})}\BibitemShut {NoStop}%
\bibitem [{\citenamefont {Lu}\ \emph {et~al.}(2015)\citenamefont {Lu},
  \citenamefont {Du}, \citenamefont {Cui},\ and\ \citenamefont
  {Zong}}]{Lu2015}%
  \BibitemOpen
  \bibfield  {author} {\bibinfo {author} {\bibfnamefont {Y.}~\bibnamefont
  {Lu}}, \bibinfo {author} {\bibfnamefont {Y.-L.}\ \bibnamefont {Du}}, \bibinfo
  {author} {\bibfnamefont {Z.-F.}\ \bibnamefont {Cui}}, \ and\ \bibinfo
  {author} {\bibfnamefont {H.-S.}\ \bibnamefont {Zong}},\ }\href {\doibase
  10.1140/epjc/s10052-015-3720-2} {\bibfield  {journal} {\bibinfo  {journal}
  {EPJC}\ }\textbf {\bibinfo {volume} {75}},\ \bibinfo {pages} {495} (\bibinfo
  {year} {2015})}\BibitemShut {NoStop}%
\bibitem [{\citenamefont {Du}\ \emph {et~al.}(2015)\citenamefont {Du},
  \citenamefont {Lu}, \citenamefont {Xu}, \citenamefont {Cui}, \citenamefont
  {Shi},\ and\ \citenamefont {Zong}}]{doi:10.1142/S0217751X15501997}%
  \BibitemOpen
  \bibfield  {author} {\bibinfo {author} {\bibfnamefont {Y.-L.}\ \bibnamefont
  {Du}}, \bibinfo {author} {\bibfnamefont {Y.}~\bibnamefont {Lu}}, \bibinfo
  {author} {\bibfnamefont {S.-S.}\ \bibnamefont {Xu}}, \bibinfo {author}
  {\bibfnamefont {Z.-F.}\ \bibnamefont {Cui}}, \bibinfo {author} {\bibfnamefont
  {C.}~\bibnamefont {Shi}}, \ and\ \bibinfo {author} {\bibfnamefont {H.-S.}\
  \bibnamefont {Zong}},\ }\href {\doibase 10.1142/S0217751X15501997} {\bibfield
   {journal} {\bibinfo  {journal} {Mod. Phys. A}\ }\textbf {\bibinfo {volume}
  {30}},\ \bibinfo {pages} {1550199} (\bibinfo {year} {2015})},\ \Eprint
  {http://arxiv.org/abs/https://doi.org/10.1142/S0217751X15501997}
  {https://doi.org/10.1142/S0217751X15501997} \BibitemShut {NoStop}%
\bibitem [{\citenamefont {Luo}(2016)}]{LUO201675}%
  \BibitemOpen
  \bibfield  {author} {\bibinfo {author} {\bibfnamefont {X.}~\bibnamefont
  {Luo}},\ }\href {\doibase https://doi.org/10.1016/j.nuclphysa.2016.03.025}
  {\bibfield  {journal} {\bibinfo  {journal} {Nucl. Phys. A}\ }\textbf
  {\bibinfo {volume} {956}},\ \bibinfo {pages} {75 } (\bibinfo {year}
  {2016})},\ \bibinfo {note} {the XXV International Conference on
  Ultrarelativistic Nucleus-Nucleus Collisions: Quark Matter 2015}\BibitemShut
  {NoStop}%
\bibitem [{\citenamefont {Gr\"af}\ \emph {et~al.}(2012)\citenamefont {Gr\"af},
  \citenamefont {Bleicher},\ and\ \citenamefont {Li}}]{PhysRevC.85.044901}%
  \BibitemOpen
  \bibfield  {author} {\bibinfo {author} {\bibfnamefont {G.}~\bibnamefont
  {Gr\"af}}, \bibinfo {author} {\bibfnamefont {M.}~\bibnamefont {Bleicher}}, \
  and\ \bibinfo {author} {\bibfnamefont {Q.}~\bibnamefont {Li}},\ }\href
  {\doibase 10.1103/PhysRevC.85.044901} {\bibfield  {journal} {\bibinfo
  {journal} {Phys. Rev. C}\ }\textbf {\bibinfo {volume} {85}},\ \bibinfo
  {pages} {044901} (\bibinfo {year} {2012})}\BibitemShut {NoStop}%
\bibitem [{\citenamefont {Bass}\ \emph {et~al.}(1998)\citenamefont {Bass},
  \citenamefont {Belkacem},\ and\ \citenamefont {et~al}}]{BASS1998255}%
  \BibitemOpen
  \bibfield  {author} {\bibinfo {author} {\bibfnamefont {S.}~\bibnamefont
  {Bass}}, \bibinfo {author} {\bibfnamefont {M.}~\bibnamefont {Belkacem}}, \
  and\ \bibinfo {author} {\bibnamefont {et~al}},\ }\href {\doibase
  https://doi.org/10.1016/S0146-6410(98)00058-1} {\bibfield  {journal}
  {\bibinfo  {journal} {Prog. Part. Nucl. Phys}\ }\textbf {\bibinfo {volume}
  {41}},\ \bibinfo {pages} {255 } (\bibinfo {year} {1998})}\BibitemShut
  {NoStop}%
\bibitem [{\citenamefont {Palhares}\ \emph {et~al.}(2011)\citenamefont
  {Palhares}, \citenamefont {Fraga},\ and\ \citenamefont
  {Kodama}}]{JPhysG.38.085101}%
  \BibitemOpen
  \bibfield  {author} {\bibinfo {author} {\bibfnamefont {L.~F.}\ \bibnamefont
  {Palhares}}, \bibinfo {author} {\bibfnamefont {E.~S.}\ \bibnamefont {Fraga}},
  \ and\ \bibinfo {author} {\bibfnamefont {T.}~\bibnamefont {Kodama}},\ }\href
  {http://stacks.iop.org/0954-3899/38/i=8/a=085101} {\bibfield  {journal}
  {\bibinfo  {journal} {J. Phys. G}\ }\textbf {\bibinfo {volume} {38}},\
  \bibinfo {pages} {085101} (\bibinfo {year} {2011})}\BibitemShut {NoStop}%
\bibitem [{\citenamefont {Wang}\ \emph {et~al.}(2018)\citenamefont {Wang},
  \citenamefont {Xia},\ and\ \citenamefont
  {Zong}}]{doi:10.1142/S0217732318502322}%
  \BibitemOpen
  \bibfield  {author} {\bibinfo {author} {\bibfnamefont {Q.-W.}\ \bibnamefont
  {Wang}}, \bibinfo {author} {\bibfnamefont {Y.}~\bibnamefont {Xia}}, \ and\
  \bibinfo {author} {\bibfnamefont {H.-S.}\ \bibnamefont {Zong}},\ }\href
  {\doibase 10.1142/S0217732318502322} {\bibfield  {journal} {\bibinfo
  {journal} {Mod. Phys. Lett. A}\ }\textbf {\bibinfo {volume} {33}},\ \bibinfo
  {pages} {1850232} (\bibinfo {year} {2018})},\ \Eprint
  {http://arxiv.org/abs/https://doi.org/10.1142/S0217732318502322}
  {https://doi.org/10.1142/S0217732318502322} \BibitemShut {NoStop}%
\bibitem [{\citenamefont {Pan}\ \emph {et~al.}(2017)\citenamefont {Pan},
  \citenamefont {Cui}, \citenamefont {Chang},\ and\ \citenamefont
  {Zong}}]{IntJModPhysA.32.1750067}%
  \BibitemOpen
  \bibfield  {author} {\bibinfo {author} {\bibfnamefont {Z.}~\bibnamefont
  {Pan}}, \bibinfo {author} {\bibfnamefont {Z.-F.}\ \bibnamefont {Cui}},
  \bibinfo {author} {\bibfnamefont {C.-H.}\ \bibnamefont {Chang}}, \ and\
  \bibinfo {author} {\bibfnamefont {H.-S.}\ \bibnamefont {Zong}},\ }\href
  {\doibase 10.1142/S0217751X17500671} {\bibfield  {journal} {\bibinfo
  {journal} {Int. J. Mod. Phys. A}\ }\textbf {\bibinfo {volume} {32}},\
  \bibinfo {pages} {1750067} (\bibinfo {year} {2017})}\BibitemShut {NoStop}%
\bibitem [{\citenamefont {Bhattacharyya}\ \emph {et~al.}(2013)\citenamefont
  {Bhattacharyya}, \citenamefont {Deb}, \citenamefont {Ghosh}, \citenamefont
  {Ray},\ and\ \citenamefont {Sur}}]{PhysRevD.87.054009}%
  \BibitemOpen
  \bibfield  {author} {\bibinfo {author} {\bibfnamefont {A.}~\bibnamefont
  {Bhattacharyya}}, \bibinfo {author} {\bibfnamefont {P.}~\bibnamefont {Deb}},
  \bibinfo {author} {\bibfnamefont {S.~K.}\ \bibnamefont {Ghosh}}, \bibinfo
  {author} {\bibfnamefont {R.}~\bibnamefont {Ray}}, \ and\ \bibinfo {author}
  {\bibfnamefont {S.}~\bibnamefont {Sur}},\ }\href {\doibase
  10.1103/PhysRevD.87.054009} {\bibfield  {journal} {\bibinfo  {journal} {Phys.
  Rev. D}\ }\textbf {\bibinfo {volume} {87}},\ \bibinfo {pages} {054009}
  (\bibinfo {year} {2013})}\BibitemShut {NoStop}%
\bibitem [{\citenamefont {Tripolt}\ \emph {et~al.}(2014)\citenamefont
  {Tripolt}, \citenamefont {Braun}, \citenamefont {Klein},\ and\ \citenamefont
  {Schaefer}}]{PhysRevD.90.054012}%
  \BibitemOpen
  \bibfield  {author} {\bibinfo {author} {\bibfnamefont {R.-A.}\ \bibnamefont
  {Tripolt}}, \bibinfo {author} {\bibfnamefont {J.}~\bibnamefont {Braun}},
  \bibinfo {author} {\bibfnamefont {B.}~\bibnamefont {Klein}}, \ and\ \bibinfo
  {author} {\bibfnamefont {B.-J.}\ \bibnamefont {Schaefer}},\ }\href {\doibase
  10.1103/PhysRevD.90.054012} {\bibfield  {journal} {\bibinfo  {journal} {Phys.
  Rev. D}\ }\textbf {\bibinfo {volume} {90}},\ \bibinfo {pages} {054012}
  (\bibinfo {year} {2014})}\BibitemShut {NoStop}%
\bibitem [{\citenamefont {Shi}\ \emph {et~al.}(2018{\natexlab{a}})\citenamefont
  {Shi}, \citenamefont {Xia}, \citenamefont {Jia},\ and\ \citenamefont
  {Zong}}]{Shi:2018tsq}%
  \BibitemOpen
  \bibfield  {author} {\bibinfo {author} {\bibfnamefont {C.}~\bibnamefont
  {Shi}}, \bibinfo {author} {\bibfnamefont {Y.}~\bibnamefont {Xia}}, \bibinfo
  {author} {\bibfnamefont {W.}~\bibnamefont {Jia}}, \ and\ \bibinfo {author}
  {\bibfnamefont {H.}~\bibnamefont {Zong}},\ }\href {\doibase
  10.1007/s11433-017-9177-4} {\bibfield  {journal} {\bibinfo  {journal} {Sci.
  China Phys. Mech. Astron.}\ }\textbf {\bibinfo {volume} {61}},\ \bibinfo
  {pages} {082021} (\bibinfo {year} {2018}{\natexlab{a}})}\BibitemShut
  {NoStop}%
\bibitem [{\citenamefont {Li}\ \emph {et~al.}(2019)\citenamefont {Li},
  \citenamefont {Cui}, \citenamefont {Zhou}, \citenamefont {An}, \citenamefont
  {Zhang},\ and\ \citenamefont {Zong}}]{LI2019298}%
  \BibitemOpen
  \bibfield  {author} {\bibinfo {author} {\bibfnamefont {B.-L.}\ \bibnamefont
  {Li}}, \bibinfo {author} {\bibfnamefont {Z.-F.}\ \bibnamefont {Cui}},
  \bibinfo {author} {\bibfnamefont {B.-W.}\ \bibnamefont {Zhou}}, \bibinfo
  {author} {\bibfnamefont {S.}~\bibnamefont {An}}, \bibinfo {author}
  {\bibfnamefont {L.-P.}\ \bibnamefont {Zhang}}, \ and\ \bibinfo {author}
  {\bibfnamefont {H.-S.}\ \bibnamefont {Zong}},\ }\href {\doibase
  https://doi.org/10.1016/j.nuclphysb.2018.11.015} {\bibfield  {journal}
  {\bibinfo  {journal} {Nucl. Phys. B}\ }\textbf {\bibinfo {volume} {938}},\
  \bibinfo {pages} {298 } (\bibinfo {year} {2019})}\BibitemShut {NoStop}%
\bibitem [{\citenamefont {Shi}\ \emph {et~al.}(2018{\natexlab{b}})\citenamefont
  {Shi}, \citenamefont {Jia}, \citenamefont {Sun}, \citenamefont {Zhang},\ and\
  \citenamefont {Zong}}]{1674-1137-42-2-023101}%
  \BibitemOpen
  \bibfield  {author} {\bibinfo {author} {\bibfnamefont {C.}~\bibnamefont
  {Shi}}, \bibinfo {author} {\bibfnamefont {W.}~\bibnamefont {Jia}}, \bibinfo
  {author} {\bibfnamefont {A.}~\bibnamefont {Sun}}, \bibinfo {author}
  {\bibfnamefont {L.}~\bibnamefont {Zhang}}, \ and\ \bibinfo {author}
  {\bibfnamefont {H.}~\bibnamefont {Zong}},\ }\href
  {http://stacks.iop.org/1674-1137/42/i=2/a=023101} {\bibfield  {journal}
  {\bibinfo  {journal} {CPC}\ }\textbf {\bibinfo {volume} {42}},\ \bibinfo
  {pages} {023101} (\bibinfo {year} {2018}{\natexlab{b}})}\BibitemShut
  {NoStop}%
\bibitem [{\citenamefont {Balian}\ and\ \citenamefont
  {Bloch}(1970)}]{BALIAN1970401}%
  \BibitemOpen
  \bibfield  {author} {\bibinfo {author} {\bibfnamefont {R.}~\bibnamefont
  {Balian}}\ and\ \bibinfo {author} {\bibfnamefont {C.}~\bibnamefont {Bloch}},\
  }\href {\doibase https://doi.org/10.1016/0003-4916(70)90497-5} {\bibfield
  {journal} {\bibinfo  {journal} {Ann. Phys}\ }\textbf {\bibinfo {volume}
  {60}},\ \bibinfo {pages} {401 } (\bibinfo {year} {1970})}\BibitemShut
  {NoStop}%
\bibitem [{\citenamefont {Fischer}\ and\ \citenamefont
  {Luecker}(2013)}]{FISCHER20131036}%
  \BibitemOpen
  \bibfield  {author} {\bibinfo {author} {\bibfnamefont {C.~S.}\ \bibnamefont
  {Fischer}}\ and\ \bibinfo {author} {\bibfnamefont {J.}~\bibnamefont
  {Luecker}},\ }\href {\doibase https://doi.org/10.1016/j.physletb.2012.11.054}
  {\bibfield  {journal} {\bibinfo  {journal} {Phys. Lett. B}\ }\textbf
  {\bibinfo {volume} {718}},\ \bibinfo {pages} {1036 } (\bibinfo {year}
  {2013})}\BibitemShut {NoStop}%
\bibitem [{\citenamefont {Shi}\ \emph {et~al.}(2016)\citenamefont {Shi},
  \citenamefont {Du}, \citenamefont {Xu}, \citenamefont {Liu},\ and\
  \citenamefont {Zong}}]{PhysRevD.93.036006}%
  \BibitemOpen
  \bibfield  {author} {\bibinfo {author} {\bibfnamefont {C.}~\bibnamefont
  {Shi}}, \bibinfo {author} {\bibfnamefont {Y.-L.}\ \bibnamefont {Du}},
  \bibinfo {author} {\bibfnamefont {S.-S.}\ \bibnamefont {Xu}}, \bibinfo
  {author} {\bibfnamefont {X.-J.}\ \bibnamefont {Liu}}, \ and\ \bibinfo
  {author} {\bibfnamefont {H.-S.}\ \bibnamefont {Zong}},\ }\href {\doibase
  10.1103/PhysRevD.93.036006} {\bibfield  {journal} {\bibinfo  {journal} {Phys.
  Rev. D}\ }\textbf {\bibinfo {volume} {93}},\ \bibinfo {pages} {036006}
  (\bibinfo {year} {2016})}\BibitemShut {NoStop}%
\bibitem [{\citenamefont {Xu}\ \emph {et~al.}(2015{\natexlab{a}})\citenamefont
  {Xu}, \citenamefont {Cui}, \citenamefont {Wang}, \citenamefont {Shi},
  \citenamefont {Yang},\ and\ \citenamefont {Zong}}]{PhysRevD.91.056003}%
  \BibitemOpen
  \bibfield  {author} {\bibinfo {author} {\bibfnamefont {S.-S.}\ \bibnamefont
  {Xu}}, \bibinfo {author} {\bibfnamefont {Z.-F.}\ \bibnamefont {Cui}},
  \bibinfo {author} {\bibfnamefont {B.}~\bibnamefont {Wang}}, \bibinfo {author}
  {\bibfnamefont {Y.-M.}\ \bibnamefont {Shi}}, \bibinfo {author} {\bibfnamefont
  {Y.-C.}\ \bibnamefont {Yang}}, \ and\ \bibinfo {author} {\bibfnamefont
  {H.-S.}\ \bibnamefont {Zong}},\ }\href {\doibase 10.1103/PhysRevD.91.056003}
  {\bibfield  {journal} {\bibinfo  {journal} {Phys. Rev. D}\ }\textbf {\bibinfo
  {volume} {91}},\ \bibinfo {pages} {056003} (\bibinfo {year}
  {2015}{\natexlab{a}})}\BibitemShut {NoStop}%
\bibitem [{\citenamefont {Xu}\ \emph {et~al.}(2015{\natexlab{b}})\citenamefont
  {Xu}, \citenamefont {Chen}, \citenamefont {Clo\"et}, \citenamefont {Roberts},
  \citenamefont {Segovia},\ and\ \citenamefont {Zong}}]{PhysRevD.92.114034}%
  \BibitemOpen
  \bibfield  {author} {\bibinfo {author} {\bibfnamefont {S.-S.}\ \bibnamefont
  {Xu}}, \bibinfo {author} {\bibfnamefont {C.}~\bibnamefont {Chen}}, \bibinfo
  {author} {\bibfnamefont {I.~C.}\ \bibnamefont {Clo\"et}}, \bibinfo {author}
  {\bibfnamefont {C.~D.}\ \bibnamefont {Roberts}}, \bibinfo {author}
  {\bibfnamefont {J.}~\bibnamefont {Segovia}}, \ and\ \bibinfo {author}
  {\bibfnamefont {H.-S.}\ \bibnamefont {Zong}},\ }\href {\doibase
  10.1103/PhysRevD.92.114034} {\bibfield  {journal} {\bibinfo  {journal} {Phys.
  Rev. D}\ }\textbf {\bibinfo {volume} {92}},\ \bibinfo {pages} {114034}
  (\bibinfo {year} {2015}{\natexlab{b}})}\BibitemShut {NoStop}%
\bibitem [{\citenamefont {Li}\ \emph {et~al.}(2016)\citenamefont {Li},
  \citenamefont {Chang}, \citenamefont {Gao}, \citenamefont {Roberts},
  \citenamefont {Schmidt},\ and\ \citenamefont {Zong}}]{PhysRevD.93.114033}%
  \BibitemOpen
  \bibfield  {author} {\bibinfo {author} {\bibfnamefont {B.-L.}\ \bibnamefont
  {Li}}, \bibinfo {author} {\bibfnamefont {L.}~\bibnamefont {Chang}}, \bibinfo
  {author} {\bibfnamefont {F.}~\bibnamefont {Gao}}, \bibinfo {author}
  {\bibfnamefont {C.~D.}\ \bibnamefont {Roberts}}, \bibinfo {author}
  {\bibfnamefont {S.~M.}\ \bibnamefont {Schmidt}}, \ and\ \bibinfo {author}
  {\bibfnamefont {H.-S.}\ \bibnamefont {Zong}},\ }\href {\doibase
  10.1103/PhysRevD.93.114033} {\bibfield  {journal} {\bibinfo  {journal} {Phys.
  Rev. D}\ }\textbf {\bibinfo {volume} {93}},\ \bibinfo {pages} {114033}
  (\bibinfo {year} {2016})}\BibitemShut {NoStop}%
\bibitem [{\citenamefont {Roberts}\ and\ \citenamefont
  {Schmidt}(2000)}]{ROBERTS2000S1}%
  \BibitemOpen
  \bibfield  {author} {\bibinfo {author} {\bibfnamefont {C.}~\bibnamefont
  {Roberts}}\ and\ \bibinfo {author} {\bibfnamefont {S.}~\bibnamefont
  {Schmidt}},\ }\href {\doibase https://doi.org/10.1016/S0146-6410(00)90011-5}
  {\bibfield  {journal} {\bibinfo  {journal} {Prog. Part. Nucl. Phys}\ }\textbf
  {\bibinfo {volume} {45}},\ \bibinfo {pages} {S1 } (\bibinfo {year}
  {2000})}\BibitemShut {NoStop}%
\bibitem [{\citenamefont {Rusnak}\ and\ \citenamefont
  {Furnstahl}(1995)}]{Rusnak1995}%
  \BibitemOpen
  \bibfield  {author} {\bibinfo {author} {\bibfnamefont {J.~J.}\ \bibnamefont
  {Rusnak}}\ and\ \bibinfo {author} {\bibfnamefont {R.~J.}\ \bibnamefont
  {Furnstahl}},\ }\href {\doibase 10.1007/BF01289507} {\bibfield  {journal}
  {\bibinfo  {journal} {Z. Phys. A}\ }\textbf {\bibinfo {volume} {352}},\
  \bibinfo {pages} {345} (\bibinfo {year} {1995})}\BibitemShut {NoStop}%
\bibitem [{\citenamefont {Kiriyama}\ and\ \citenamefont
  {Hosaka}(2003)}]{PhysRevD.67.085010}%
  \BibitemOpen
  \bibfield  {author} {\bibinfo {author} {\bibfnamefont {O.}~\bibnamefont
  {Kiriyama}}\ and\ \bibinfo {author} {\bibfnamefont {A.}~\bibnamefont
  {Hosaka}},\ }\href {\doibase 10.1103/PhysRevD.67.085010} {\bibfield
  {journal} {\bibinfo  {journal} {Phys. Rev. D}\ }\textbf {\bibinfo {volume}
  {67}},\ \bibinfo {pages} {085010} (\bibinfo {year} {2003})}\BibitemShut
  {NoStop}%
\bibitem [{\citenamefont {Kiriyama}(2005)}]{PhysRevD.72.054009}%
  \BibitemOpen
  \bibfield  {author} {\bibinfo {author} {\bibfnamefont {O.}~\bibnamefont
  {Kiriyama}},\ }\href {\doibase 10.1103/PhysRevD.72.054009} {\bibfield
  {journal} {\bibinfo  {journal} {Phys. Rev. D}\ }\textbf {\bibinfo {volume}
  {72}},\ \bibinfo {pages} {054009} (\bibinfo {year} {2005})}\BibitemShut
  {NoStop}%
\bibitem [{\citenamefont {Madsen}(1994)}]{PhysRevD.50.3328}%
  \BibitemOpen
  \bibfield  {author} {\bibinfo {author} {\bibfnamefont {J.}~\bibnamefont
  {Madsen}},\ }\href {\doibase 10.1103/PhysRevD.50.3328} {\bibfield  {journal}
  {\bibinfo  {journal} {Phys. Rev. D}\ }\textbf {\bibinfo {volume} {50}},\
  \bibinfo {pages} {3328} (\bibinfo {year} {1994})}\BibitemShut {NoStop}%
\bibitem [{\citenamefont {Lugones}\ \emph {et~al.}(2013)\citenamefont
  {Lugones}, \citenamefont {Grunfeld},\ and\ \citenamefont
  {Ajmi}}]{PhysRevC.88.045803}%
  \BibitemOpen
  \bibfield  {author} {\bibinfo {author} {\bibfnamefont {G.}~\bibnamefont
  {Lugones}}, \bibinfo {author} {\bibfnamefont {A.~G.}\ \bibnamefont
  {Grunfeld}}, \ and\ \bibinfo {author} {\bibfnamefont {M.~A.}\ \bibnamefont
  {Ajmi}},\ }\href {\doibase 10.1103/PhysRevC.88.045803} {\bibfield  {journal}
  {\bibinfo  {journal} {Phys. Rev. C}\ }\textbf {\bibinfo {volume} {88}},\
  \bibinfo {pages} {045803} (\bibinfo {year} {2013})}\BibitemShut {NoStop}%
\bibitem [{\citenamefont {Neergaard}\ and\ \citenamefont
  {Madsen}(1999)}]{PhysRevD.60.054011}%
  \BibitemOpen
  \bibfield  {author} {\bibinfo {author} {\bibfnamefont {G.}~\bibnamefont
  {Neergaard}}\ and\ \bibinfo {author} {\bibfnamefont {J.}~\bibnamefont
  {Madsen}},\ }\href {\doibase 10.1103/PhysRevD.60.054011} {\bibfield
  {journal} {\bibinfo  {journal} {Phys. Rev. D}\ }\textbf {\bibinfo {volume}
  {60}},\ \bibinfo {pages} {054011} (\bibinfo {year} {1999})}\BibitemShut
  {NoStop}%
\bibitem [{\citenamefont {Gao}\ and\ \citenamefont
  {Liu}(2016{\natexlab{a}})}]{PhysRevD.94.076009}%
  \BibitemOpen
  \bibfield  {author} {\bibinfo {author} {\bibfnamefont {F.}~\bibnamefont
  {Gao}}\ and\ \bibinfo {author} {\bibfnamefont {Y.-x.}\ \bibnamefont {Liu}},\
  }\href {\doibase 10.1103/PhysRevD.94.076009} {\bibfield  {journal} {\bibinfo
  {journal} {Phys. Rev. D}\ }\textbf {\bibinfo {volume} {94}},\ \bibinfo
  {pages} {076009} (\bibinfo {year} {2016}{\natexlab{a}})}\BibitemShut
  {NoStop}%
\bibitem [{\citenamefont {Gao}\ and\ \citenamefont
  {Liu}(2016{\natexlab{b}})}]{PhysRevD.94.094030}%
  \BibitemOpen
  \bibfield  {author} {\bibinfo {author} {\bibfnamefont {F.}~\bibnamefont
  {Gao}}\ and\ \bibinfo {author} {\bibfnamefont {Y.-x.}\ \bibnamefont {Liu}},\
  }\href {\doibase 10.1103/PhysRevD.94.094030} {\bibfield  {journal} {\bibinfo
  {journal} {Phys. Rev. D}\ }\textbf {\bibinfo {volume} {94}},\ \bibinfo
  {pages} {094030} (\bibinfo {year} {2016}{\natexlab{b}})}\BibitemShut
  {NoStop}%
\bibitem [{\citenamefont {Maris}\ and\ \citenamefont
  {Tandy}(1999)}]{PhysRevC.60.055214}%
  \BibitemOpen
  \bibfield  {author} {\bibinfo {author} {\bibfnamefont {P.}~\bibnamefont
  {Maris}}\ and\ \bibinfo {author} {\bibfnamefont {P.~C.}\ \bibnamefont
  {Tandy}},\ }\href {\doibase 10.1103/PhysRevC.60.055214} {\bibfield  {journal}
  {\bibinfo  {journal} {Phys. Rev. C}\ }\textbf {\bibinfo {volume} {60}},\
  \bibinfo {pages} {055214} (\bibinfo {year} {1999})}\BibitemShut {NoStop}%
\bibitem [{\citenamefont {Maris}\ and\ \citenamefont
  {Roberts}(1997)}]{PhysRevC.56.3369}%
  \BibitemOpen
  \bibfield  {author} {\bibinfo {author} {\bibfnamefont {P.}~\bibnamefont
  {Maris}}\ and\ \bibinfo {author} {\bibfnamefont {C.~D.}\ \bibnamefont
  {Roberts}},\ }\href {\doibase 10.1103/PhysRevC.56.3369} {\bibfield  {journal}
  {\bibinfo  {journal} {Phys. Rev. C}\ }\textbf {\bibinfo {volume} {56}},\
  \bibinfo {pages} {3369} (\bibinfo {year} {1997})}\BibitemShut {NoStop}%
\bibitem [{\citenamefont {Qin}\ \emph {et~al.}(2011)\citenamefont {Qin},
  \citenamefont {Chang}, \citenamefont {Chen}, \citenamefont {Liu},\ and\
  \citenamefont {Roberts}}]{PhysRevLett.106.172301}%
  \BibitemOpen
  \bibfield  {author} {\bibinfo {author} {\bibfnamefont {S.-x.}\ \bibnamefont
  {Qin}}, \bibinfo {author} {\bibfnamefont {L.}~\bibnamefont {Chang}}, \bibinfo
  {author} {\bibfnamefont {H.}~\bibnamefont {Chen}}, \bibinfo {author}
  {\bibfnamefont {Y.-x.}\ \bibnamefont {Liu}}, \ and\ \bibinfo {author}
  {\bibfnamefont {C.~D.}\ \bibnamefont {Roberts}},\ }\href {\doibase
  10.1103/PhysRevLett.106.172301} {\bibfield  {journal} {\bibinfo  {journal}
  {Phys. Rev. Lett.}\ }\textbf {\bibinfo {volume} {106}},\ \bibinfo {pages}
  {172301} (\bibinfo {year} {2011})}\BibitemShut {NoStop}%
\bibitem [{\citenamefont {H\"oll}\ \emph {et~al.}(1999)\citenamefont {H\"oll},
  \citenamefont {Maris},\ and\ \citenamefont {Roberts}}]{PhysRevC.59.1751}%
  \BibitemOpen
  \bibfield  {author} {\bibinfo {author} {\bibfnamefont {A.}~\bibnamefont
  {H\"oll}}, \bibinfo {author} {\bibfnamefont {P.}~\bibnamefont {Maris}}, \
  and\ \bibinfo {author} {\bibfnamefont {C.~D.}\ \bibnamefont {Roberts}},\
  }\href {\doibase 10.1103/PhysRevC.59.1751} {\bibfield  {journal} {\bibinfo
  {journal} {Phys. Rev. C}\ }\textbf {\bibinfo {volume} {59}},\ \bibinfo
  {pages} {1751} (\bibinfo {year} {1999})}\BibitemShut {NoStop}%
\bibitem [{\citenamefont {Grunfeld}\ and\ \citenamefont
  {Lugones}(2018)}]{Grunfeld2018}%
  \BibitemOpen
  \bibfield  {author} {\bibinfo {author} {\bibfnamefont {A.~G.}\ \bibnamefont
  {Grunfeld}}\ and\ \bibinfo {author} {\bibfnamefont {G.}~\bibnamefont
  {Lugones}},\ }\href@noop {} {\bibfield  {journal} {\bibinfo  {journal} {Eur.
  Phys. J. C}\ }\textbf {\bibinfo {volume} {78}},\ \bibinfo {pages} {640}
  (\bibinfo {year} {2018})}\BibitemShut {NoStop}%
\bibitem [{\citenamefont {Ball}\ and\ \citenamefont
  {Chiu}(1980)}]{PhysRevD.22.2542}%
  \BibitemOpen
  \bibfield  {author} {\bibinfo {author} {\bibfnamefont {J.~S.}\ \bibnamefont
  {Ball}}\ and\ \bibinfo {author} {\bibfnamefont {T.-W.}\ \bibnamefont
  {Chiu}},\ }\href {\doibase 10.1103/PhysRevD.22.2542} {\bibfield  {journal}
  {\bibinfo  {journal} {Phys. Rev. D}\ }\textbf {\bibinfo {volume} {22}},\
  \bibinfo {pages} {2542} (\bibinfo {year} {1980})}\BibitemShut {NoStop}%
\bibitem [{\citenamefont {Chang}\ \emph {et~al.}(2011)\citenamefont {Chang},
  \citenamefont {Liu},\ and\ \citenamefont {Roberts}}]{PhysRevLett.106.072001}%
  \BibitemOpen
  \bibfield  {author} {\bibinfo {author} {\bibfnamefont {L.}~\bibnamefont
  {Chang}}, \bibinfo {author} {\bibfnamefont {Y.-X.}\ \bibnamefont {Liu}}, \
  and\ \bibinfo {author} {\bibfnamefont {C.~D.}\ \bibnamefont {Roberts}},\
  }\href {\doibase 10.1103/PhysRevLett.106.072001} {\bibfield  {journal}
  {\bibinfo  {journal} {Phys. Rev. Lett.}\ }\textbf {\bibinfo {volume} {106}},\
  \bibinfo {pages} {072001} (\bibinfo {year} {2011})}\BibitemShut {NoStop}%
\end{thebibliography}%
\end{document}